\documentclass[9pt,twocolumn,twoside]{pnas-new}

\templatetype{pnasresearcharticle} 

\usepackage{tabularx}
\usepackage{booktabs}
\renewcommand{\arraystretch}{1.15}
\newcolumntype{Y}{>{\raggedright\arraybackslash}X}

\newcommand{\secref}[1]{Section~\ref{#1}}
\newcommand{\subsecref}[2]{Section~\ref{#1}\ref{#2}}

\newcommand{\figref}[1]{Fig.~\ref{#1}}
\newcommand{\tabref}[1]{Table~\ref{#1}}

\newcommand{\logit}{\ensuremath{\operatorname{logit}}}

\usepackage[binary-units=true,detect-weight=true,detect-family=true]{siunitx}
\title{Machine-Guided Discovery of a Real-World Rogue Wave Model}

\author[a,b,1]{Dion Häfner}
\author[c]{Johannes Gemmrich} 
\author[b]{Markus Jochum}

\affil[a]{Pasteur Labs, Brooklyn, NY, USA}
\affil[b]{Niels Bohr Institute, University of Copenhagen, Copenhagen, Denmark}
\affil[c]{University of Victoria, Victoria, British Columbia, Canada}

\leadauthor{Häfner} 

\significancestatement{Machine learning has had a transformative impact on predictive science and engineering. But due to their black-box nature, better machine learning models do not always lead to greater human understanding, the first goal of science. We show how this can be overcome by using machine learning to transform a vast database of wave observations into a new, human-readable equation for the occurrence probability of rogue waves --- rare ocean waves that routinely damage ships and offshore structures. This equation can be analyzed and incorporated into the research canon. Our work demonstrates the potential of causal analysis, machine learning, and symbolic regression to drive scientific discovery in a real-world application.}

\authorcontributions{All authors conceived the project. D.H. drafted, implemented, and executed the analysis. All authors interpreted
the results. D.H. drafted the manuscript. All authors reviewed the manuscript}
\authordeclaration{Dion Häfner’s work has been partially funded by the Danish Offshore Technology Centre (DOTC).
Johannes Gemmrich and Markus Jochum declare no potential competing interests.}
\correspondingauthor{\textsuperscript{1}To whom correspondence should be addressed. E-mail: dion.haefner@simulation.science}


\begin{abstract}
    Big data and large-scale machine learning have had a profound impact on science and engineering, particularly in fields focused on forecasting and prediction. Yet, it is still not clear how we can use the superior pattern matching abilities of machine learning models for scientific \emph{discovery}. This is because the goals of machine learning and science are generally not aligned. In addition to being accurate, scientific theories must also be causally consistent with the underlying physical process and allow for human analysis, reasoning, and manipulation to advance the field.
    In this paper, we present a case study on discovering a new symbolic model for oceanic rogue waves from data using causal analysis, deep learning, parsimony-guided model selection, and symbolic regression.
    We train an artificial neural network on causal features from an extensive dataset of observations from wave buoys, while selecting for predictive performance and causal invariance. We apply symbolic regression to distill this black-box model into a mathematical equation that retains the neural network's predictive capabilities, while allowing for interpretation in the context of existing wave theory. The resulting model reproduces known behavior, generates well-calibrated probabilities, and achieves better predictive scores on unseen data than current theory. This showcases how machine learning can facilitate inductive scientific discovery, and paves the way for more accurate rogue wave forecasting.
\end{abstract}

\dates{This manuscript was compiled on \today}
\doi{\textbf{Postprint}\hspace{7pt}|\hspace{7pt}Published version @ \url{https://www.pnas.org/doi/10.1073/pnas.2306275120}}

\begin{document}

\maketitle
\thispagestyle{firststyle}
\ifthenelse{\boolean{shortarticle}}{\ifthenelse{\boolean{singlecolumn}}{\abscontentformatted}{\abscontent}}{}

\begin{figure*}[tbh]
    \centering
    \includegraphics[width=\textwidth]{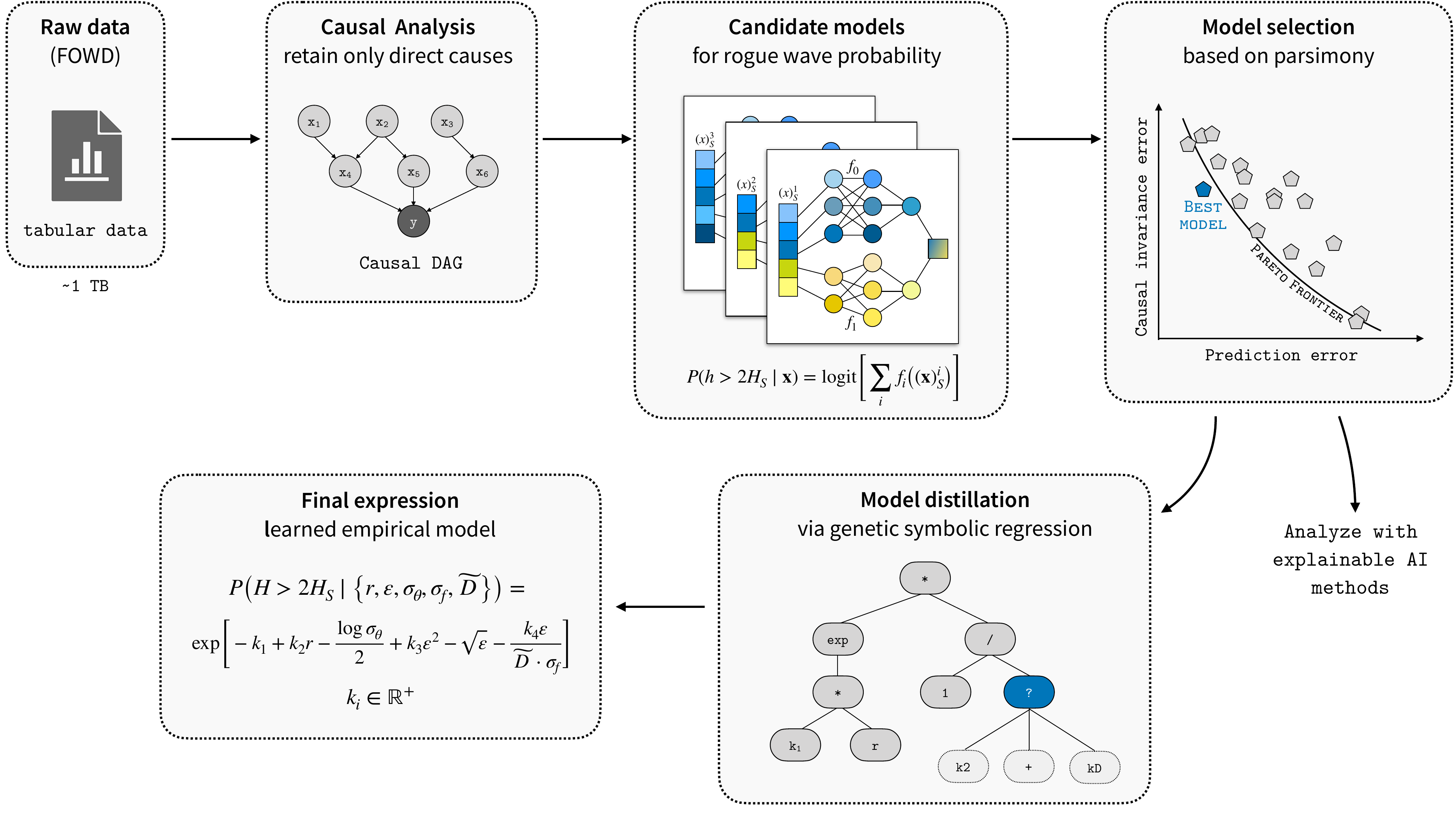}
    \caption{Overview of our study. Starting out with large amounts of tabular data from wave buoys, we use a causal analysis to identify the most important features for predicting rogue waves. We then train an ensemble of neural networks on subsets of these features, and select the best one based on its predictive performance and causal invariance. Finally, we use symbolic regression to distill the model into a concise mathematical equation. We analyze the neural network and symbolic expression in terms of their performance on unseen data, and compare them to existing theory. This closes the arc between data, machine learning, and theory.}
    \label{fig:abstract}
\end{figure*}

\dropcap{R}ogue waves are extreme ocean waves that have caused countless accidents, often with fatal consequences \citep{didenkulova_catalogue_2019}. They are defined as waves whose crest-to-trough height $H$ exceeds a threshold relative to the significant wave height $H_s$. The significant wave height is defined as four times the standard deviation of the sea surface elevation. Here, we use a rogue wave criterion with a threshold of 2.0:

\begin{equation}
    H / H_s > 2.0
\end{equation}

\noindent A rogue wave is therefore by definition an unlikely sample from the tail of the wave height distribution, and can in principle occur by chance under any circumstance. This makes them difficult to analyze, and requires massive amounts of data. Therefore, research has mostly focused on theory and idealized experiments in wave tanks, often considering only 1-dimensional wave propagation \citep{dudley_rogue_2019}. However, the availability of large observation arrays \citep{behrens_cdip_2019} makes them an ideal target for machine-learning based analysis \citep{hafner_real-world_2021,cattrell2018can}.

In this study, we present a neural network-based model that predicts rogue wave probabilities from the sea state, trained solely on observations from buoys \citep{hafner_fowd_2021}. The resulting model respects the causal structure of rogue wave generation; therefore, it can generalize to unseen physical regimes, is robust to distributional shift, and can be used to infer the relative importance of rogue wave generation mechanisms. 

While a causally consistent neural network is useful for prediction and qualitative insight into the physical dynamics, the ability for scientists to analyze, test, and manipulate a model is crucial to recognize its limitations and integrate it into the research canon. Despite advances in interpretable AI \citep{molnar_interpretable_nodate}, this is still a major challenge for most machine learning models.

To address this, we transform our neural network into a concise equation using symbolic regression \citep{cranmer2020discovering,cranmer2023interpretable}. The resulting model combines several known wave dynamics, outperforms current theory in predicting rogue wave occurrences, and can be interpreted within the context of wave theory. We see this as an example of ``data-mining inspired induction'' \citep{voit2019perspective}, an extension to the scientific method in which machine learning guides the discovery of new scientific theories.

We achieve this through the following recipe (\figref{fig:abstract}):

\begin{enumerate}
    \item A-priori analysis of causal pathways that leads to a set of presumed causal parameters (\secref{sec:causes}).
    \item Training an ensemble of regularized neural network predictors, and parsimony-guided model selection based on causal invariance (\secref{sec:model}).
    \item Distillation of the neural network into a concise mathematical expression via symbolic regression (\secref{sec:symreg}).
\end{enumerate}

\noindent Finally, we analyze both the neural network and symbolic model in the context of current wave theory (\secref{sec:results}). Both models reproduce well-known behavior and point towards new insights regarding the relative importance of different mechanisms in the real ocean.

\section{A causal graph for rogue wave generation} \label{sec:causes}

\begin{figure*}
    \centering
    \includegraphics[width=\textwidth]{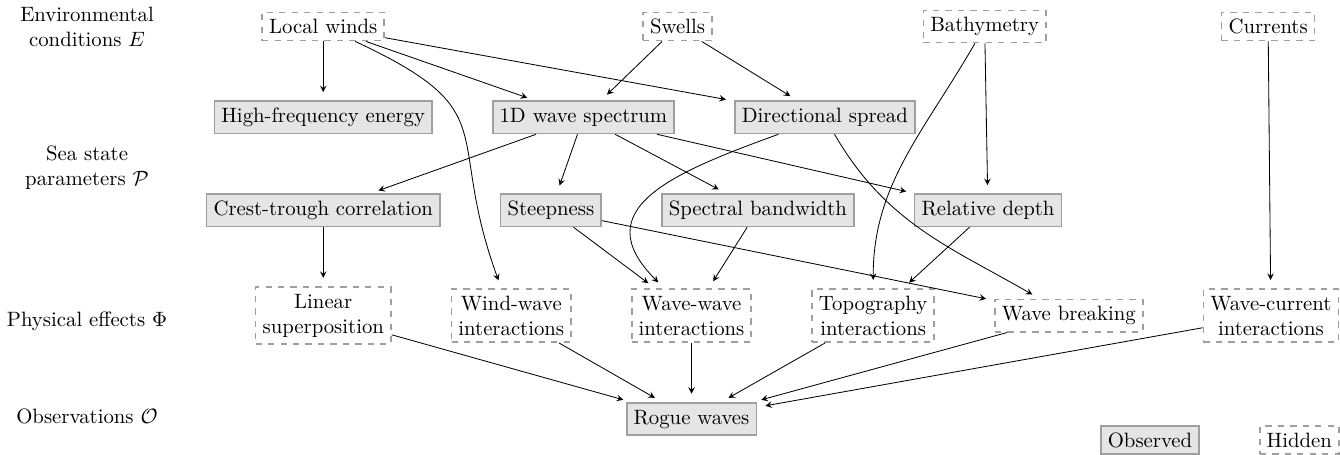}
    \caption{The causes of rogue waves as a causal DAG (directed acyclic graph). Arrows $A \to B$ imply that $A$ causes $B$.}
    \label{fig:dag}
\end{figure*}

To create a causal machine learning model it is crucial to expose it only to parameters with causal relevance. Otherwise, the model may prefer to encode spurious associations over true causal relationships, simply because they can be easier to learn. This requires us to identify the causal structure of rogue wave generation.

There are several hypothesized causes of rogue waves \citep[see][for an overview]{adcock_physics_2014}. Typically, research focuses on linear superposition in finite-bandwidth seas \citep{tayfun_wave-height_2007}, wave breaking \citep{miche_mouvements_1944}, and wave-wave interactions in weakly nonlinear seas \citep{gemmrich_dynamical_2011,fedele_real_2016} or through the modulational instability \citep{onorato_extreme_2006}. Apart from these universal mechanisms, there are also countless possible interactions with localized features such as non-uniform topography \citep{trulsen_laboratory_2012}, wave-current interactions like in the Agulhas \citep{mallory_abnormal_1974} or the Antarctic Circumpolar Current \citep{didenkulova_rogue_2021}, or crossing sea states at high crossing angles affecting wave breaking \citep{mcallister_laboratory_2019}. We call this set of mechanisms the \emph{physical effects} $\Phi$.

Since ocean waves are generated by a complex dynamical system, their true cause is a set of extrinsic \emph{environmental conditions} $E$ that are high-dimensional and not feasible to capture in full detail. However, most physical effects are mediated by one or several \emph{sea state parameters} $\mathcal{P}$, which are the characteristic aggregated parameters that appear in theoretical models of the respective wave dynamics, and that are included in operational wave forecasts. In this study, we would like obtain a model that relates relevant sea state parameters $\mathcal{P}$ to wave observations $\mathcal{O}$, which ideally also lets us infer the relative importance of physical effects $\Phi$. 

The go-to tool to analyze causal relationships is a causal DAG \citep[Directed Acyclic Graph;][]{pearl2009causality}. In a causal DAG, nodes represent variables and edges $A \to B$ imply that $A$ is a cause of $B$ (usually in the probabilistic sense in that the probability distribution $P(B)$ depends on $A$). 

We create a causal graph for rogue wave formation based on the hypothesized causal mechanisms discussed above and their corresponding theoretical models and parameters (\figref{fig:dag}).
Following this causal structure, we use the following set of sea state parameters as candidates for representing the various causal pathways (see Methods section for more information on each parameter):

\begin{itemize}
    \item Crest-trough correlation $r$, to account for the linear effect of wave groups on crest-to-trough rogue waves \citep{tayfun1990distribution}. $r$ is the dominant causal factor behind linear rogue wave formation \citep{hafner_real-world_2021}.
    \item Steepness $\varepsilon$ governing weakly nonlinear effects, such as second-order and third-order bound waves, and wave breaking \citep{miche_mouvements_1944,goda_reanalysis_2010}.
    \item Relative high-frequency energy $E_h$ (fraction of total energy contained in the spectral band \SIrange{0.25}{1.5}{\hertz}) as a proxy for the strength of local winds \citep{tang2022impact}.
    \item Relative depth $\widetilde{D}$ (based on peak wavelength), which is central for nonlinear shallow-water effects \citep{korteweg1895xli,janssen_shallow-water_2018} and wave breaking \citep{miche_mouvements_1944}.
    \item Dominant directional spread $\sigma_\theta$, which has an influence on third-order nonlinear waves \citep{janssen_shallow-water_2018} and wave breaking \citep{mcallister_laboratory_2019}.
    \item Spectral bandwidth $\nu_f$ (narrowness) and $\sigma_f$ (peakedness), appearing for example in the expression for the influence of third-order nonlinear waves \citep{janssen_shallow-water_2018}.
\end{itemize}

\noindent We also include a number of derived parameters that commonly appear in wave models and govern certain nonlinear (wave-wave) phenomena:

\begin{itemize}
    \item Benjamin-Feir index $\mathrm{BFI}$, which controls third-order nonlinear free waves \citep{janssen_shallow-water_2018} and the modulational instability \citep{janssen_nonlinear_2003}.
    \item Ursell number $\mathrm{Ur}$, which quantifies nonlinear effects in shallow water \citep{ursell_long-wave_1953}.
    \item Directionality index $R$ (the ratio of directional spread and spectral bandwidth), which has an influence on third-order nonlinear free waves and is typically used in conjunction with the BFI \citep{janssen_shallow-water_2018}.
\end{itemize}

\noindent These parameters cover most causal pathways towards rogue wave generation. Still, there are some at least partially unobserved causes, as we do not have access to data on local winds, topography, or currents. Additionally, our in-situ measurements are potentially biased estimates of the true sea state parameters, and there is no guarantee that any given training procedure will converge to the true causal model. This implies that we cannot rely on a model being causally consistent by design; instead, we perform a-posteriori verification on the learned models to find the perfect trade-off between causal consistency and predictive performance (see \secref{sec:model}\ref{sec:icp}).

\section{An approximately causal neural network} \label{sec:model}
\subsection{Input data} \label{sec:data}

We use the Free Ocean Wave Dataset \citep[FOWD,][]{hafner_fowd_2021}, which contains 1.4 billion wave measurements recorded by the 158 CDIP wave buoys \citep{behrens_cdip_2019} along the Pacific and Atlantic coasts of the US, Hawaii, and overseas US territories. Water depths range between \SIrange{10}{4000}{\metre}, and we require a significant wave height of at least \SI{1}{\metre}. Each buoy records the sea surface elevation at a sampling frequency of \SI{1.28}{\hertz}, producing over \num{700} years of time series in total. FOWD extracts every zero-crossing wave from the surface elevation data and computes a number of characteristic sea state parameters from the history of the wave within a sliding window.

Due to the massive data volume of the full FOWD catalogue ($\sim\SI{1}{\tera\byte}$), we use an aggregated version that maps each sea state to the maximum wave height of the following 100 waves \citep[as in][]{hafner_real-world_2021}. This reduces the data volume by a factor of 100 and inflates all rogue wave probabilities to a bigger value $\hat{p}$. We correct for this via $p = 1 - (1 - \hat{p})^{1/100}$, assuming that rogue waves occur independently from each other. This is a good approximation in most conditions, but may underestimate seas with a strong group structure (see \secref{sec:discussion}\ref{sec:limitations}).

The final dataset has 12.9M data points containing over \num{100000} rogue waves exceeding 2 times the significant wave height. Our dataset is freely available for download (see Data Availability section).

\subsection{Neural network architecture}

\begin{figure*}[tbh]
    \centering
    \includegraphics[width=.7\textwidth]{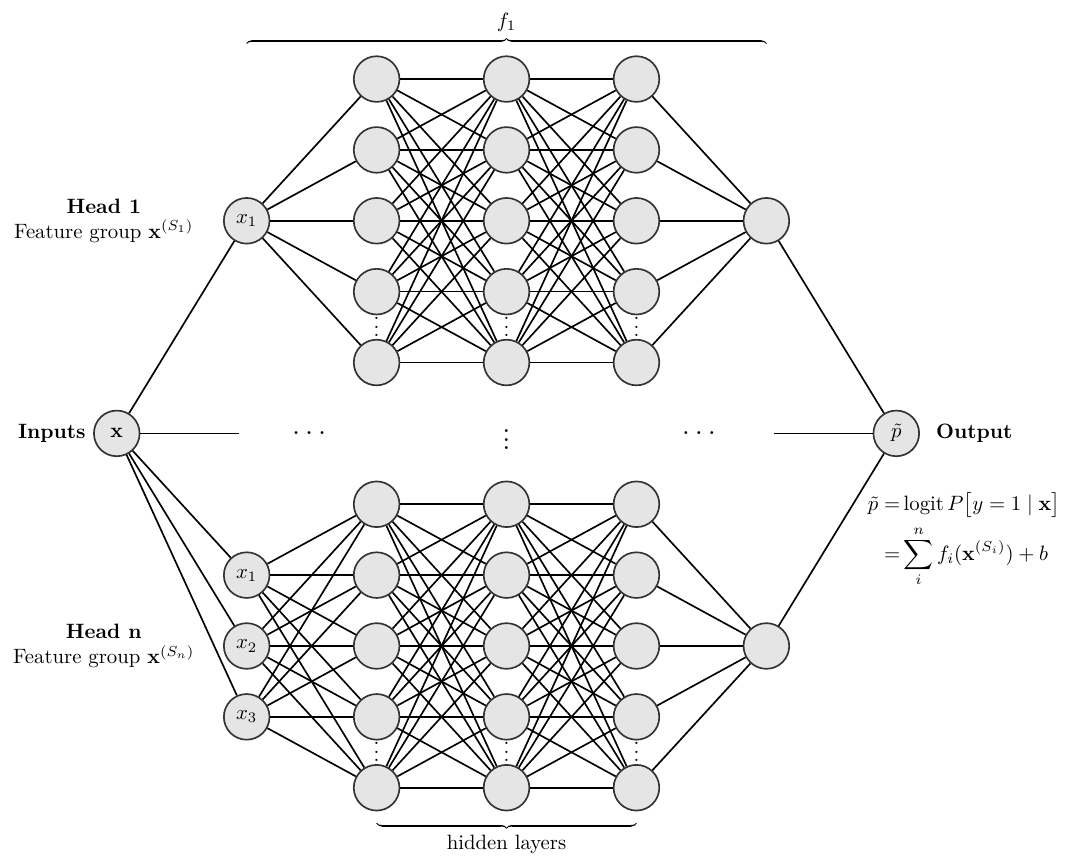}
    \caption{Neural network architecture (multi-head FCN) used to predict rogue wave probabilities. Each input head receives a different subset of the full parameter set $\mathbf{x}$ to limit the amount of non-causal interactions between parameters.}
    \label{fig:nn}
\end{figure*}

The probability to measure a rogue wave based on the sea state can be modelled as a sum of nonlinear functions, each of which only depends on a subset of the sea state parameters representing a different causal path (act via different \emph{physical effects} in \figref{fig:dag}):

\begin{equation}
    \logit P\big(y = 1 \;\big\vert\; \textbf{x}\big) \sim \sum_i f_i\big(\mathbf{x}^{(S_i)}\big) + b \label{eq:decomp}
\end{equation}

\noindent Here, $y$ is a binary label indicating whether the current wave is a rogue wave, $\mathbf{x}^{(S_i)}$ is the i-th subset of all causal sea state parameters $\mathbf{x}$, $\logit(p) = \log (p) - \log(1 - p)$ is the logit function, $f_i$ are arbitrary nonlinear functions to be learned, and $b$ is a constant bias term. 

By including only a subset $\mathbf{x}^{(S_i)}$ of all parameters $\mathbf{x}$ as input for $f_i$, we can restrict which parameters may interact non-additively with each other, which is an additional regularizing constraint that increases interpretability and prevents interactions between inputs from different causal pathways. For example, to include the effects of linear superposition and nonlinear corrections for free and bound waves \citep[as in][]{ecmwf_part_2021}, \eqref{eq:decomp} can be written as:

\begin{equation}
    \logit P\big(y = 1 \;\big\vert\; \textbf{x}\big) \sim \underbrace{f_1(r)}_{\text{linear}} + \underbrace{f_2(\mathrm{BFI}, R)}_{\text{free waves}} + \underbrace{f_3(\varepsilon, \widetilde{D})}_{\text{bound waves}}
\end{equation}

\noindent We use a neural network with fully-connected layers (FCN) to model the functions $f_i$, which are universal function approximators \citep{hornik_approximation_1991}, and that can be trained efficiently for large amounts of data. The set of functions $f_i$ can be represented as a single multi-head FCN with a linear output layer (\figref{fig:nn}). We use a small feed-forward architecture with 3 hidden layers and ReLU activation functions \citep[rectified linear units,][]{nair_rectified_2010}. To the best of our knowledge, this is the first time that a multi-head neural network has been used to restrict the interactions between input parameters to be consistent with a causal model.

The neural network outputs a scalar $\tilde{p} \in (-\infty, \infty)$, the log-odds of a rogue wave occurrence for the given sea state. For training, we use the Adam optimizer \citep{kingma2014adam} and backpropagation to minimize a cross-entropy loss for binary classification with an added $\ell_2$ regularization term for kernel parameters:

\begin{equation}
    L(p, y, \theta) = y \cdot \log (p) + (1-y) \cdot \log (1-p) + \lambda \lVert \theta \rVert_2
\end{equation}

\noindent with predicted probability $p = \logit^{-1}(\tilde{p})$, observed labels $y \in \{0, 1\}$ (rogue wave or not), and neural network kernel parameters $\theta$.

To estimate uncertainties in the neural network parameters and resulting predictions, we use Gaussian stochastic weight averaging \citep[SWAG,][]{maddox_simple_2019}. For this, we train the network for 50 epochs, then start recording the optimizer trajectory after each epoch for another 50 epochs. The observed covariance structure of the sampled parameters is used to construct a multivariate Gaussian approximation of the loss surface that we can sample from. This results in slightly better predictions, and gives us a way to quantify how confident the neural network is in its predictions.

\subsection{Causal consistency and predictive accuracy} \label{sec:icp}

Although we include only input parameters that we assume to have a direct causal connection with rogue wave generation, there is no guarantee that the neural network will infer the correct causal model. In fact, the presence of measurement bias and unobserved causal paths makes it unlikely that the model will converge to the true causal structure. To search for an approximately causally consistent model we will have to quantify its causal performance.

We achieve this through the concept of invariant causal prediction \citep[ICP;][]{peters_causal_2016,peters_elements_2017}. The key insight behind ICP is that the parameters of the true causal model will be invariant under distributional shift, that is, an intervention on an upstream ``environment'' node in the causal graph that controls which distribution the data is drawn from. Re-training the model on data with different spurious correlations \emph{between} features should still lead to the same dependency of the target \emph{on} the features \citep[see also][]{heinze2018invariant}.

We split the dataset randomly into separate training and validation sets, in chunks of 1M waves. We train the model on the full training dataset and perform ICP on the validation dataset, which we partition into subsets representing different conditions in space, time, depth, spectral properties, and degrees of non-linearity (\tabref{tab:subsets}). This changes the dominant characteristics of the waves in each subset (representing e.g.\ storm and swell conditions), inducing distributional shift. Then, we re-train the model separately on each subset and compute the root-mean-square difference between predictions of the re-trained model $P_k$ and the full model $P_\text{tot}$ on the $k$-th data subset $\mathbf{x}_{(k)}$:

\begin{table}
    \caption{The subsets of the validation dataset used to evaluate model performance and invariance.} \label{tab:subsets}
    \centering
    




\footnotesize
\begin{tabularx}{0.98\linewidth}{lYr}
    {Subset name} & {Condition} & {\# waves} \\ \midrule
    southern-california & Longitude $\in (-123.5, -117)\si{\degree}$, latitude $\in (32, 38)\si{\degree}$ & 265M \\
    deep-stations & Water depth $> \SI{1000}{\metre}$ & 28M \\
    shallow-stations & Water depth $< \SI{100}{\metre}$ & 154M \\[1.2ex]
    summer & Day of year $\in (160, 220)$ & 51M \\
    winter & Day of year $\in (0, 60)$ & 91M \\
    Hs $>$ 3m & $H_s > \SI{3}{\metre}$ & 58M \\[1.2ex]
    high-frequency & Relative swell energy $<0.15$ & 43M \\
    low-frequency & Relative swell energy $>0.7$ & 46M \\
    long-period & Mean zero-crossing period $> \SI{9}{\second}$ & 100M \\[1.2ex]
    short-period & Mean zero-crossing period $< \SI{6}{\second}$ & 42M \\
    cnoidal & Ursell number $> 8$ & 40M \\
    weakly-nonlinear & Steepness $> 0.04$ & 83M \\[1.2ex]
    low-spread & Directional spread $< \SI{20}{\degree}$ & 25M \\
    high-spread & Directional spread $> \SI{40}{\degree}$ & 25M \\[1.2ex]
    full & (all validation data) & 472M \\
    \bottomrule
\end{tabularx}

\end{table}

\begin{align}
    \mathcal{E}_k^2 = \frac{1}{n_k} \sum_i^{n_k} \bigg( \logit P_k\big(\mathbf{x}^{(k)}_i \big) - \logit P_\text{tot} \big(\mathbf{x}^{(k)}_i \big) \bigg)^2 \label{eq:icp-env}
\end{align}

\noindent As the total consistency error we use the root-mean-square of \eqref{eq:icp-env} across all environments:

\begin{equation}
    \mathcal{E} = \sqrt{\frac{1}{n_E} \sum_k^{n_E} \mathcal{E}_k^2}
\end{equation}

\noindent Under a noise-free, infinite dataset and an unbiased training process that always identifies the true causal model we would find $\mathcal{E} = 0$, i.e., re-training the model on the unseen data subset would not contribute any new information and leave the model perfectly invariant. Since all of these assumptions are violated here, we merely search for an approximately causal model that minimizes $\mathcal{E}$.

However, we cannot use $\mathcal{E}$ as the only criterion when selecting a model. The invariance error can only account for change in the prediction (variance), but not for its overall closeness to the true solution (bias). Therefore, we select a model that is Pareto-optimal with respect to the invariance error $\mathcal{E}$ and a predictive score $\mathcal{L}$. This will not establish absolute causal consistency, but will allow us to select a model that is near-optimal given the constraints.

For $\mathcal{L}$ we use the log of the likelihood ratio between the predictions of our neural network and a baseline model that predicts the empirical base rate $\overline{y}_k = \frac{1}{n} \sum_i^n y_{k,i}$, averaged over all environments $k$:

\begin{align}
    \mathcal{L}({p}, \overline{y}) &= \frac{1}{n_E} \sum_k^{n_E} \big( I(p_k) - I(\overline{y}_k) \big) \\
    I(x) &= x \cdot \log (x) + (1-x) \cdot \log (1-x)
\end{align}

\noindent To evaluate model calibration (the tendency to produce over- or under-confident probabilities), we compute a calibration curve by binning the predicted rogue wave probabilities. We then compare each bin to the observed rogue wave frequency, and compute the weighted root-mean-square residual between measured ($\overline{y}_i$) and predicted ($p_i$) log-odds:

\begin{equation}
    \mathcal{C} = \sqrt{\sum_{i=1}^{n_b} w_i \big( \logit (p_i) - \logit (\overline{y}_i) \big)^2} \label{eq:calibration}
\end{equation}

\noindent To account for uncertainty in the observations (e.g.\ close to the extremes), the weights $w_k$ are based on the \SI{33}{\percent} credible interval of $\overline{y}_i \sim \operatorname{Beta}(n^+_i, n^-_i)$ with $n^+_i$ rogue and $n^-_i$ non-rogue measurements. This is similar to the expected calibration error \citep{xenopoulos2022calibrate}, but models data uncertainty directly. We use a uniform bin size (in logit space) of 0.1.

\subsection{Model selection}
\label{sec:selection}

\begin{figure*}
    \centering
    \includegraphics[width=.7\linewidth]{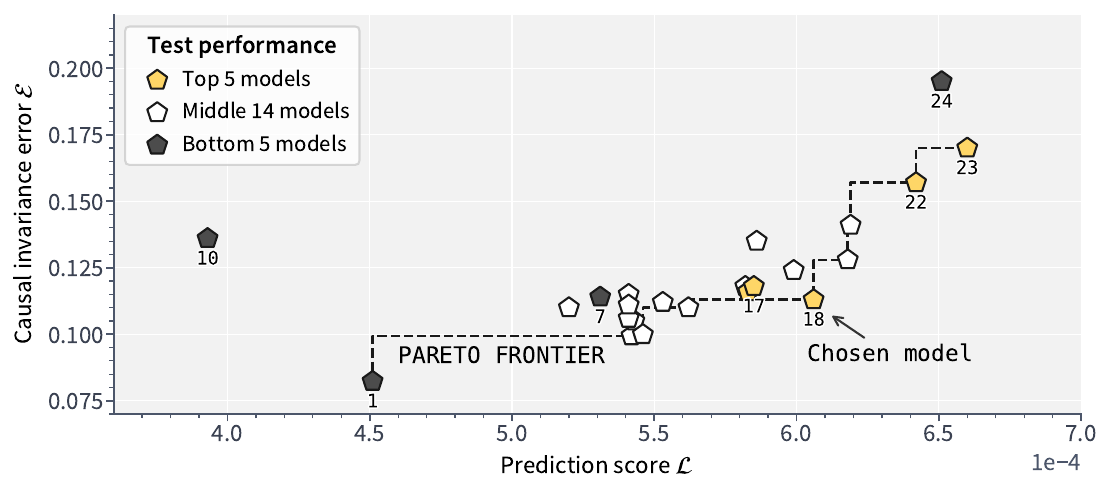}
    \caption{There is a clear trade-off between causal invariance ($\mathcal{E}$) and predictive performance ($\mathcal{L}$) of our neural network predictors. We choose the model that lies in the most convex part of the Pareto frontier. Scores are evaluated on validation data. Test performance is based on prediction scores on held-out test data (from unseen stations).} \label{fig:pareto}
\end{figure*}

We train a total of 24 candidate models on different subsets of the relevant causal parameters (as identified in \secref{sec:causes}) and varying number of input heads (between 1 and 3). We evaluate their performance in terms of calibration, predictive performance, and causal consistency (\tabref{tab:experiments}).

\begin{figure}
    \centering
    \includegraphics[width=.8\linewidth]{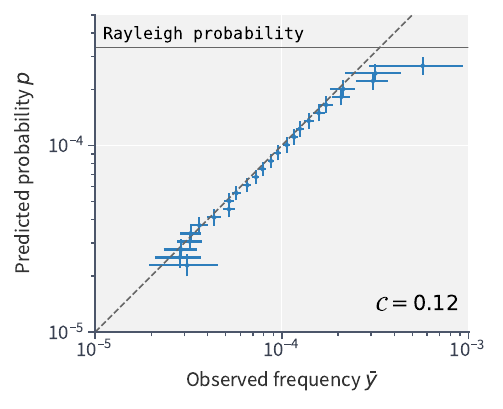}
    \caption{Our model outputs well-calibrated probabilities, even for unseen stations. Shown is the binned predicted probability $p$ vs.\ the observed rogue wave frequency $\overline{y}$ on the test data. Error bars for $p$ indicate 3 standard deviations estimated via SWAG sampling. Error bars for $\overline{y}$ indicate \SI{95}{\percent} credible interval assuming $\overline{y}_i \sim \operatorname{Beta}(n^+_i, n^-_i)$. Bins with less than 10 observed rogue waves are excluded. Dashed line indicates perfect calibration. Solid line indicates probability as predicted by linear theory in the narrow-bandwidth limit \citep[Rayleigh distribution;][]{longuet1952statisticaldistribution}.} \label{fig:calibration}
\end{figure}

We observe a clear anti-correlation between model complexity and predictive score on one hand and causal consistency on the other hand (\figref{fig:pareto}). This is evidence that more complex models are indeed less biased but exploit more non-causal connections. We perform model selection based on \emph{parsimony}: A good model is one where a small increase in either predictive performance or causal consistency implies a large decrease in the other, i.e., where the Pareto front is convex. This is similar to the metric used by PySR \citep{cranmer2023interpretable} to select the best symbolic regression model (\secref{sec:symreg}).

Based on this, we choose model 18 with parameter groups $S_1 = \{r\}, S_2 = \{ \varepsilon, \sigma_\theta, \sigma_f, \widetilde{D} \}$ (i.e., a model with two input heads) as the reference model for further analysis. The chosen model produces well-calibrated probabilities (\figref{fig:calibration}), and is among the 5 best models in terms of predictive performance on the test dataset (not used during training or selection), despite using only 5 features with at most 4-way interactions.

The relatively low number of input features allows us to analyze the model in detail using explainable AI methods (Section \ref{sec:results}\ref{sec:nn-analysis}).

\section{Learning an empirical equation for rogue wave risk} \label{sec:symreg}

To make our model fully interpretable, we transform the learned neural network into an equation via symbolic regression.
Common approaches to symbolic regression include Eureqa \citep{schmidt2009distilling}, AI Feynman \citep{udrescu2020ai}, SINDy \citep{brunton2016discovering}, and QLattice \citep{brolos2021approach}. Here, we use PySR \citep{cranmer2023interpretable,cranmer2020discovering}, a symbolic regression package based on genetic programming \citep{holland1992genetic}. Genetic algorithms build a large ensemble of candidate models and select the best ones, before mutating and recombining them into the next generation. In the case of symbolic regression, mathematical expressions are represented as a tree of constants and elementary symbols. In principle, this allows PySR to discover expressions of unbounded complexity.

PySR's central metric to quantify the goodness of an equation is again based on \emph{parsimony}, in the form of the derivative of predictive performance with respect to the model complexity --- if the true model has been discovered, any additional complexity can at best lead to minor performance gains (by overfitting to noise in the data).

In our case, we seek to find an expression $f$ from the space of possible expression graphs $\mathcal{T}_O$ with allowed operators $O$ that approximates the rogue wave log-probability as predicted by the neural network $\mathcal{N}$ over the dataset $x$:

\begin{equation*}
    \text{Find $f \in \mathcal{T}_O$ that minimizes} ~ \sum_i \frac{1}{\operatorname{Var}(y_i)} \bigg[ f(x_i)^2 - \sigma (\mathbb{E}[y_i])^2 \bigg]
\end{equation*}

\noindent where $\sigma(x) = -\log(1 + \exp(-x))$, and $y_i$ is the set of SWAG samples from $\mathcal{N}(x_i)$. A sensible set of operators $O$ is key to ensure interpretability of the resulting expression; we choose the symbols $O=\{+, -, \times, \div, \log, \cdot^{-1}, \sqrt{\;}, \cdot^2\}$ to facilitate expressions that are similar to current theoretical models of the form $P \sim A \exp(B)$. We normalize all input features to approximately unit scale by converting directional spread to radians.

PySR assembles a league of candidate expression and presents the Pareto-optimal solutions of increasing complexity to the user. We select the best solution by hand, picking the expression with the best parsimony score that contains all input features and at least two terms containing the steepness $\varepsilon$ (to account for the various causal pathways in which steepness affects rogue waves). The final equation is shown in \figref{fig:symreg-eq}, and discussed in \secref{sec:results}\ref{sec:results-symreg}.

\section{Results} \label{sec:results}

\subsection{Neural network} \label{sec:nn-analysis}

We analyze the behavior of our neural network predictor, which reveals important insights about the physical dynamics of rogue waves and their prediction.

\subsubsection{Rogue wave models should account for crest-trough correlation, steepness, relative depth, and directionality}

Only this parameter combination achieves good causal consistency and predictive scores at the same time, and experiments that exclude any of these parameters perform unconditionally worse in either metric. Especially the exclusion of crest-trough correlation leads to catastrophic results, even when including other bandwidth measures like $\sigma_\theta$ in its place (\tabref{tab:experiments}).

This suggests that the above set of parameters represents the dominant rogue wave generation processes in the form of linear superposition in finite-bandwidth seas with a directional contribution and weakly nonlinear corrections. 

The crest-trough correlation $r$ is still lacking mainstream adoption as a rogue wave indicator \citep[for example, it is not part of ECMWF's operational forecast;][]{ecmwf_part_2021}, despite being a key parameter for crest-to-trough rogue waves \citep{tayfun1990distribution,fedele_nonlinear_2009,hafner_real-world_2021}. The other parameters are consistent with other empirical studies such as Fedele \citep{fedele_large_2019}, which considers the same parameters in conjunction with rogue crests during storms.
They are also similar to the ingredients to ECMWF's rogue wave forecast \citep{ecmwf_part_2021}, which is based on the effects of second and third-order bound and free waves and uses steepness, relative depth, directional spread, and spectral bandwidth. However, in our model these parameters are combined differently; a model enforcing the same interactions (steepness and relative depth for bound wave contribution, BFI and directionality index for free wave contribution) performs poorly.

Numerous previous studies have found the BFI to be a poor predictor of rogue wave risk in realistic sea states \citep{fedele_real_2016,fedele_large_2019,gramstad_influence_2007,xiao_rogue_2013,hafner_real-world_2021,gemmrich_dynamical_2011,gemmrich2017observations} due to its strong underlying assumptions such as unidirectionality. This study extends this to the fully nonparametric and nonlinear case.

\begin{figure*}[bt]
    \centering
    \includegraphics[width=.7\textwidth]{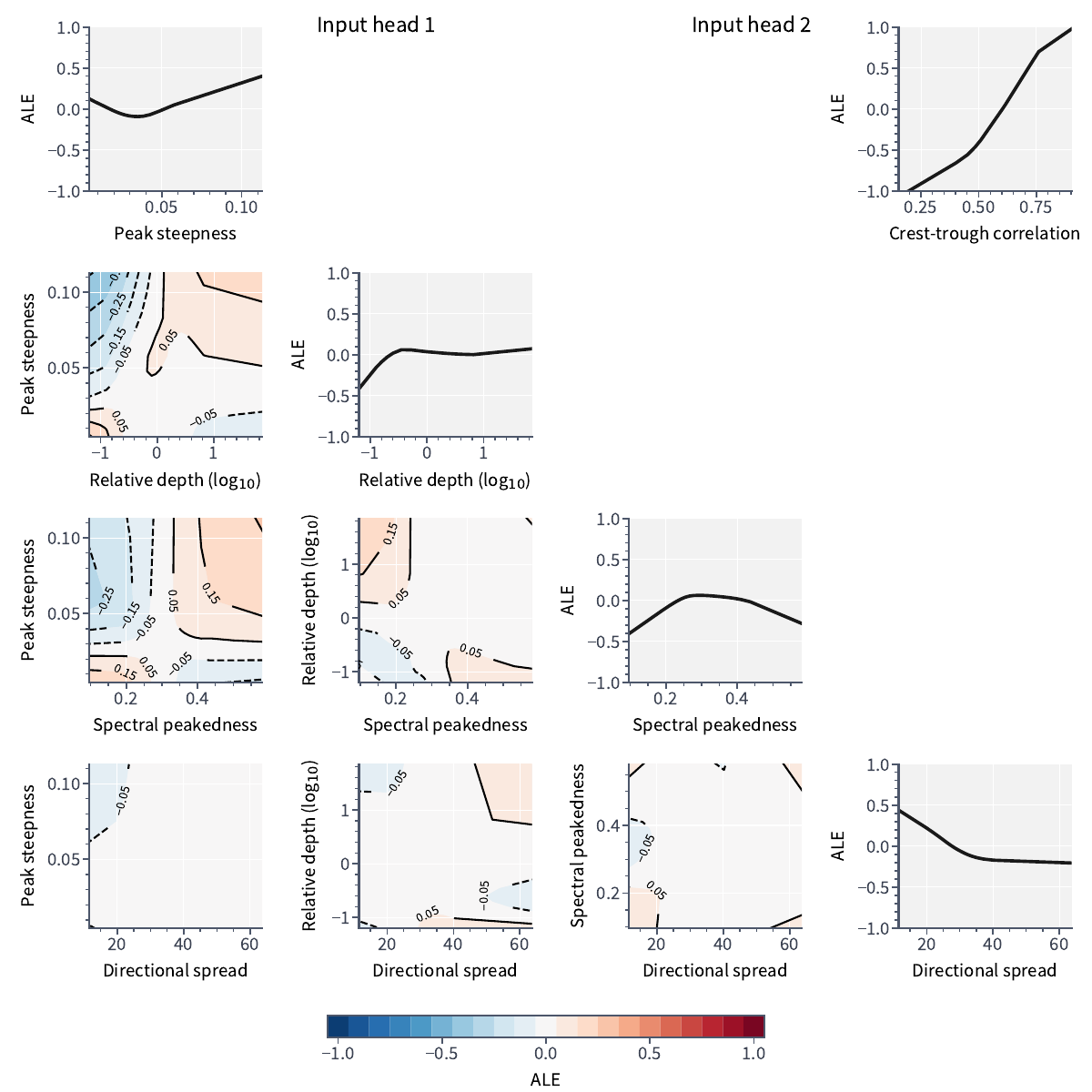}
    \caption{ALE (accumulated local effects) plot matrix for experiment 18. Shown is the change in rogue wave risk (in logits) from the average as each parameter is varied. The total effect is the sum of all 1D, 2D, and higher-order contributions (not shown).} \label{fig:ale}
\end{figure*}

We study how our model uses different parameters by visualizing their impact on the prediction of the respective head of the neural network. For this, we make use of the accumulated local effects decomposition \citep[ALE,][]{apley_visualizing_2019}, which measures the influence of infinitesimal changes in each parameter on the prediction outcome \citep[see also][]{molnar_interpretable_nodate}.
From the ALE plot (\figref{fig:ale}), we find that crest-trough correlation has by far the biggest influence of all parameters and explains about 1 order of magnitude in rogue wave risk variation, which is consistent with earlier model-free approaches \citep{hafner_real-world_2021}. To first order, higher crest-trough correlation, lower directional spread, larger relative depth (deep water), and higher steepness lead to larger rogue wave risk, but parameter interactions can lead to more complicated, non-monotonic relationships (for example in very shallow water, see \subsecref{sec:results}{sec:depth}).

\subsubsection{The Rayleigh distribution is an upper bound for real-world rogue wave risk}

Despite the clear enhancement by weakly nonlinear corrections, the Rayleigh wave height distribution remains an upper bound for real-world (crest-to-trough) rogue waves. The Rayleigh distribution is the theoretical wave height distribution for linear narrow-band waves \citep{longuet1952statisticaldistribution}, i.e., the limit $r \to 1$, $\varepsilon \to 0$, $\sigma_f \to 0$, $\widetilde{D} \to \infty$, and $\sigma_\theta \to 0$, and reads:

\begin{equation}
    P(H/H_s > k) = \exp(-2 k^2)
\end{equation}

\noindent Only in the most extreme conditions does our model predict a similarly high probability, for example for $\sigma_\theta=\SI{13}{\degree}$, $\varepsilon=0.008$, $\sigma_f=0.14$, $r=0.88$, and $\widetilde{D} =0.6$, which gives the same probability as the Rayleigh distribution, $p =\num{3.3e-4}$.

In the opposite extreme, rogue wave probabilities can fall to as little as \num{e-5} for low values of $r$ and high values of $\sigma_\theta$ (such as in a sea with a strong high-frequency component and high directional spread). This suggests that bandwidth effects can create sea states that efficiently suppress extremes.

\subsubsection{There is a clear separation between deep water and shallow water regimes} \label{sec:depth}

\begin{figure}
    \centering
    \includegraphics[width=.8\linewidth]{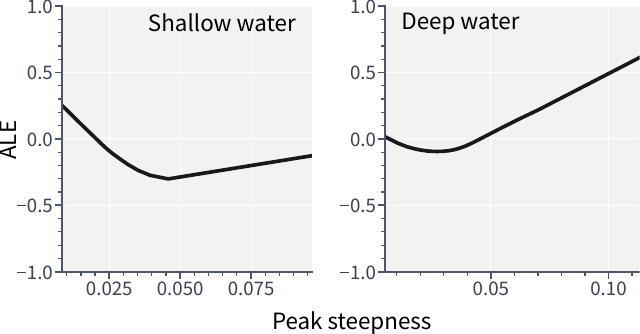}
    \caption{Our model predicts a positive association between steepness and rogue waves in deep water, and a negative association in shallow water. Shown is the 1-dimensional ALE (accumulated local effects) plot in both cases. Here, deep water are sea states with $\widetilde{D} > 3$ and shallow water with $\widetilde{D} < 0.1$.} \label{fig:steepness-regimes}
\end{figure}

All models with high causal invariance scores include an interaction between steepness and relative water depth. Looking at this more closely, we find that a stratification on deep and shallow water sea states reveals 2 distinct regimes (\figref{fig:steepness-regimes}).

In deep water, rogue wave risk is strongly positively associated with steepness, as expected from the contribution of second and third-order nonlinear bound waves \citep{janssen_shallow-water_2018}.
The opposite is true in shallow water ($\widetilde{D} < 0.1$), where we find a clear \emph{negative} association with steepness. This is likely due to depth-induced wave breaking \citep{goda_reanalysis_2010}. In very shallow waters, more sea states have a steepness close to the breaking threshold, which removes taller waves that tend to have a higher steepness than average.

\subsection{Symbolic expression} \label{sec:results-symreg}

\begin{figure*}
    \centering
    \includegraphics{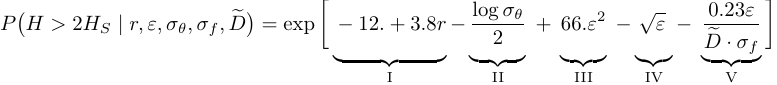}
    \caption{Our empirical equation for rogue wave risk, as identified through the distillation of our neural network predictor via symbolic regression. This equation outperforms existing wave theory on unseen stations from our dataset, while being fully interpretable. Numbered terms are discussed in \secref{sec:results}\ref{sec:results-symreg}. All floating point coefficients are rounded to two significant digits.} \label{fig:symreg-eq}
\end{figure*}

The final expression for the rogue wave probability, as discovered via symbolic regression, is given in \figref{fig:symreg-eq}. It consists of an exponential containing five additive terms:

\begin{enumerate}
    \item[(I)] $-12 + 3.8r$. The term with the largest coefficients is the one containing $r$, as expected. Comparison with the exponential term in the Tayfun distribution $P_t$, \eqref{eq:tayfun}, reveals that this is approximately a linear expansion around $r\approx 1$: 
    \begin{align}
        \log P_t(H/H_s > h) &\sim -\frac{4h^2}{1 + r} \\&= \left. -12 + 4r + \mathcal{O}(r^2) \, \right|_{r\approx 1}^{h=2}
    \end{align}

    \noindent This is an important sanity check for the model, since it shows that it is able to re-discover existing theory purely from data.
    \item[(II)] $-\log\sigma_\theta/2$. This encodes the observed enhancement for narrow sea states and has no direct relation to existing quantitative theory. Its functional form is somewhat problematic, since it causes the model to diverge for $\sigma_\theta \to 0$ (unidirectional seas). However, the model has only seen real-world seas with $\sigma_\theta \gtrsim 0.2$, so we may replace this term with one that yields similar predictions for the relevant range of $\sigma_\theta$, and does not diverge for $\sigma_\theta \to 0$.
    
    One possible candidate is 
    \begin{equation}\frac{1 - \sigma_\theta}{1 + \sigma_\theta},\end{equation}
    \noindent which has a relative RMS error of about \SI{5}{\percent} over the range $\sigma_\theta \in (20, 90)\si{\degree}$ compared to the original term.
    \item[(III)] $66\varepsilon^2$. Encodes the influence of weakly nonlinear effects for large values of $\varepsilon \gtrsim 0.1$.
    \item[(IV)] $-\sqrt{\varepsilon}$. This term encodes the observed negative association between steepness and rogue waves for low values of $\varepsilon$ that could be due to wave breaking, or may be an artifact of our sensor.
    \item[(V)] $0.23 \varepsilon / (\widetilde{D} \cdot \sigma_f)$. Since $\widetilde{D} \sim k_p D$ and $\varepsilon \sim k_p H_s$, this term is proportional to the relative wave height $\eta = H_s / D$ and $1/\sigma_f$. $\eta$ is the most important parameter in the theory of shallow-water waves, and appears for example in the Korteweg-de Vries equation \citep{korteweg1895xli}. Accordingly, this term dominates the dynamics in very shallow water. Dependencies on $1/\sigma_f$ occur in current theory \cite{janssen_shallow-water_2018}, but are usually paired with $\sigma_\theta$ to form the directionality index $R$. This suggest that term V may be incomplete, and missing physical dynamics that are not prevalent in the data.
\end{enumerate}

\noindent Overall, the equation is able to reproduce the same qualitative behavior as observed from the neural network, with the same well-calibrated outputs ($\mathcal{C}=0.14$) and predictive performance (\secref{sec:discussion}\ref{sec:validation}) on the test data.

\section{Discussion} \label{sec:discussion}

\begin{figure*}
    \centering
    \includegraphics[width=\linewidth]{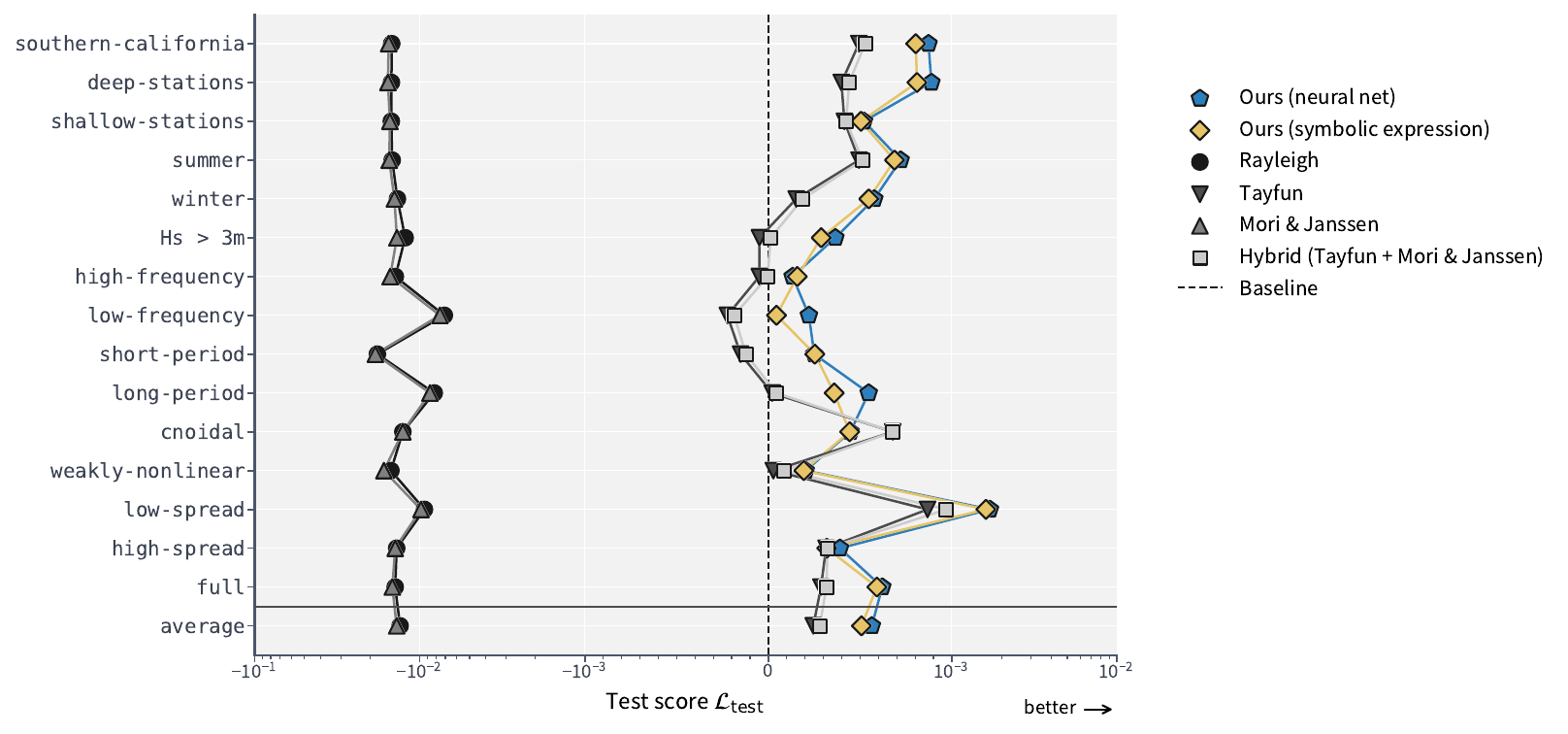}
    \caption{Comparison between our models and existing theory on held-out test data. Our models perform similar to each other and outperform existing theory on this dataset in all but one data subset (cnoidal seas). $x$-scale is linear in $(\num{-e-3}, \num{e-3})$, and logarithmic otherwise.} \label{fig:validation}
\end{figure*}

\subsection{Validation against theory} \label{sec:validation}

We test our models (neural network and symbolic equation) against existing wave theory based on their mean predictive score $\mathcal{L}$ across the environments from \tabref{tab:subsets} on the held-out test data (unseen stations). As theoretical baselines we use the models from Longuet-Higgins \citep[Rayleigh,][]{longuet1952statisticaldistribution}, Tayfun \citep{tayfun1990distribution}, Mori \& Janssen \citep{mori2011estimation}, and a hybrid combining Tayfun and Mori \& Janssen (see Methods section).

The results are shown in \figref{fig:validation}. Since the Rayleigh and Mori \& Janssen models do not account for crest-trough correlation, their predictions vastly overestimate the occurrence rate of observed rogue waves. The Tayfun and hybrid models perform better, but are still outperformed by our models except in cnoidal seas. Our models are better predictors than the baseline (predicting the empirical per-environment rogue wave frequency) in all environments.

The neural network performs better than the symbolic equation in all environments, albeit only by a small margin. This shows that the symbolic equation is able to capture the main features of the full model, despite its compact representation.

\subsection{Limitations} \label{sec:limitations}

Using only wave buoy observations for our analysis, we acknowledge the following limitations:

\begin{itemize}
\item We did not have sufficient data on local winds, currents, or topography, which implies that some relevant causal pathways are unobserved (see \figref{fig:dag}). While we expect these effects to play a minor role in bulk analysis, they could dramatically affect local rogue wave probabilities in specific conditions, for example over sloping topography \citep{trulsen_laboratory_2012} or in strong currents \citep{ying_linear_2011}.

\item We only have one-dimensional (time series) data and cannot capture imported parameters, such as solitons generated elsewhere that travel into the observation area. While we expect this to play a minor role, it could underestimate the importance of nonlinear free waves.

\item Systematic sensor bias is common in buoys and can lead to spurious causal relationships. This may obscure the true causal structure and hurt model generalization to other sensors. However, this adaptation to sensor characteristics may be desirable in forecasting scenarios, where it allows the model to synthesize several noisy quantities into more robust ones.

\item By aggregating individual waves into 100-wave chunks, we underestimate the per-wave rogue wave probability in sea states in which rogue waves do not occur independently of each other, such as seas with a strong group structure.
\end{itemize}

\noindent These limitations could potentially reduce our model's ability to detect relevant causal pathways and underestimate the true rogue wave risk. Our analysis is agnostic to the data source and can be repeated on different sources to validate our findings.

\section{Next steps}

\subsection{An improved rogue wave forecast}

Our empirical model can be compared directly to existing rogue wave risk indicators by evaluating them on forecast sea state parameters. ECMWF's operational rogue wave forecast \citep{ecmwf_part_2021} focuses on envelope wave heights which does not account for crest-trough correlation, and is conceptually similar to the Mori \& Janssen model in \secref{sec:discussion}\ref{sec:validation}. Therefore, we are confident that substantial improvements are within reach in terms of predicting crest-to-trough rogue waves, even without using a black-box model.

\subsection{Predicting super-rogue waves}

Observed wave height distributions often show a flattening of the wave height distribution towards the extreme tail \citep{gemmrich_dynamical_2011,casasprat_short-term_2010,adcock_physics_2014}. Therefore, we expect rogue wave probabilities to be more pronounced for even more extreme waves \citep[for example with $H/H_s > 2.4$, as recently observed in][]{gemmrich2022generation}.

The lack of sufficient direct observations in these regimes calls for a different strategy. One approach could be to transform this classification problem (rogue wave or not) into a regression, where the predicted variables are the free parameters of a candidate wave height probability distribution (such as shape and scale parameters of a Weibull distribution). Then, a similar analysis as in this study could be conducted for these parameters, which may reveal the main mechanisms influencing the risk for truly exceptional waves, and whether this flattening can be confirmed in our dataset.

\subsection{Commoditization of data-mining based induction}

There is a pronounced lack of established methods for machine learning aimed at scientific discovery. We have shown that incorporating and enforcing causal structure can overcome many of the shortcomings of standard machine learning approaches, like poorly calibrated predictions, non-interpretability, and incompatibility with existing theory. However, the methods we leveraged are still in their infancy and rely on further community efforts to be end-to-end automated and adopted at scale. Particularly, parsimony-based model selection (as in \secref{sec:model}\ref{sec:selection} and \secref{sec:symreg}) is still a manual process that requires a firm understanding of model intrinsics and the domain at hand. Nonetheless, we believe that the potential benefits of causal and parsimony-guided machine learning for real-world problems are too great to ignore, and we hope that this study will inspire further research in this direction.

\matmethods{
\subsection*{Sea state parameters}

Here, we give the definition of the sea state parameters used in this study. For a more thorough description of how parameters are computed from buoy displacement time series see H\"afner et al. \citep{hafner_fowd_2021}.

All parameters can be derived from the non-directional wave spectrum $\mathcal{S}(f)$, with the exception of directional spread $\sigma_\theta$, which is estimated from the horizontal motion of the buoy and taken from the raw CDIP data.

Most parameters are computed from moments of the wave spectrum, where the $n$-th moment $m_n$ is defined as

\begin{equation}
    m_n = \int_0^\infty f^n \mathcal{S}(f)\;\mathrm{d}f
\end{equation}

\noindent The expressions for the relevant sea state parameters are:

\begin{itemize}
    \item Significant wave height: \begin{equation}H_s = 4 \sqrt{m_0}\end{equation}
    \item Spectral bandwidth (narrowness): 
    \begin{equation}\nu_f = \sqrt{m_2 m_0 / m_1^2 - 1}\end{equation}
    \item Spectral bandwidth (peakedness): 
    \begin{equation}\sigma_f = \frac{m_0^2}{2 \sqrt\pi} \bigg( \int_0^\infty f \cdot \mathcal{S}(f)\;\mathrm{d}f \bigg)^{-1}\end{equation}
    \item Peak wavenumber $k_p$, computed via the peak period \citep[as in][]{young_determination_1995}:

    \begin{equation}
        \overline{T}_p = \frac{\int \mathcal{S}(f)^4 \;\mathrm{d}f}{\int f \cdot \mathcal{S}(f)^4 \;\mathrm{d}f}
    \end{equation}

    \noindent This leads to the peak wavenumber through the dispersion relation for linear waves in intermediate water of depth $D$:

    \begin{equation}
        f(k)^2 = \frac{gk}{(2 \pi)^2} \tanh(kD)
    \end{equation}

    \noindent An approximate inverse is given in Fenton \citep{fenton_numerical_1988}.

    \item Relative depth, based on the wave length $\lambda$: 
    \begin{equation} 
    \widetilde{D} = \frac{D}{\lambda} = \frac{1}{2\pi} k_p D 
    \end{equation}

    \item Peak steepness: \begin{equation}\varepsilon = H_s k_p\end{equation}

    \item Benjamin-Feir index:

    \begin{equation}
        \mathrm{BFI} = \frac{\varepsilon \nu}{\sigma_f} \sqrt{\max\{\beta / \alpha, 0\}}
    \end{equation}

    \noindent where $\nu$, $\alpha$, $\beta$ are coefficients depending only on $\widetilde{D}$ \citep[full expression given in][]{serio_computation_2005}.

    \item Directionality index:

    \begin{equation}
        R = \frac{\sigma_\theta^2}{2 \nu_f^2}
    \end{equation}

    \item Crest-trough correlation:

    \begin{gather}
        r = \frac{1}{m_0} \sqrt{\rho^2 + \lambda^2} \\
        \rho = \int_0^\infty \mathcal{S}(\omega) \cos\bigg(\omega \frac{\overline{T}}{2}\bigg) \;\mathrm{d}\omega \\
        \lambda = \int_0^\infty \mathcal{S}(\omega) \sin\bigg(\omega \frac{\overline{T}}{2}\bigg) \;\mathrm{d}\omega
        \label{eq:ctt}
    \end{gather}

    \noindent where $\omega$ is the angular frequency and $\overline{T} = m_0 / m_1$ the spectral mean period \citep{tayfun_wave-height_2007}.

\end{itemize}

\subsection*{Model implementation and hyperparameters}

All performance critical model code is implemented in JAX \citep{jax2018github}, using neural network modules from flax \citep{flax2020github} and optimizers from optax \citep{optax2020github}. We run each experiment on a single Tesla P100 GPU in about 40 minutes, including SWAG sampling and re-training on every validation subset. The whole training process can also be executed on CPU in about 2 hours.
The hyperparameters for all experiments are shown in \tabref{tab:hyperparams}.

\begin{table}[h]
    \centering
    \caption{Hyperparameters used in experiments.} \label{tab:hyperparams}
    




\begin{tabular}{ll}
    \multicolumn{2}{l}{Hyperparameters} \\ \toprule
    Optimizer & Adam \\
    Learning rate & \num{e-4} \\
    Number of hidden layers & 3 \\
    Neurons in hidden layers & $(32 / \sqrt{n_h}, 16 / \sqrt{n_h}, 8 / \sqrt{n_h})$ \\
    $\ell_2$ penalty $\lambda_2$ & \num{e-5} \\
    Number of training epochs & 50 \\
    Number of SWAG epochs & 50 \\
    Number of SWAG posterior samples & 100 \\
    Train-validation split & 60\% -- 40\% \\ \bottomrule
\end{tabular}

    \addtabletext{$n_h$: number of input heads.}
\end{table}

\subsection*{Full list of experiments}

See \tabref{tab:experiments}.

\begin{table*}
    \centering
    \caption{Full list of experiments. $\mathcal{L}$: Prediction score (higher is better). $\mathcal{E}$: Invariance error (lower is better). $\mathcal{C}$: Calibration error (lower is better). Color coding ranges between $(\text{median} - \text{IQR}, \text{median} + \text{IQR})$ with inter-quartile range IQR.} \label{tab:experiments}
    
\definecolor{C0}{RGB}{244, 109, 67}
\definecolor{C1}{RGB}{253, 174, 97}
\definecolor{C2}{RGB}{254, 224, 139}
\definecolor{C3}{RGB}{217, 239, 139}
\definecolor{C4}{RGB}{166, 217, 106}
\definecolor{C5}{RGB}{102, 189, 99}
\definecolor{Clow}{RGB}{215, 48, 39}
\definecolor{Chigh}{RGB}{26, 152, 80}

\begingroup
\renewcommand{\arraystretch}{1.2}
\begin{tabular}{rlllrrr}
    & \multicolumn{3}{c}{Feature groups} & \multicolumn{3}{c}{Scores} \\ \cmidrule(lr){2-4} \cmidrule(lr){5-7}
    ID & 1 & 2 & 3 & $\mathcal{L} \times 10^4$ & $\mathcal{E} \times 10^2$ & $\mathcal{C} \times 10^2$ \\ \midrule
    1 & $\{r\}$ &   &   & \cellcolor{Clow} \num{4.51} & \cellcolor{Chigh} \num{8.23} & \cellcolor{Chigh} \num{3.35}\\
2 & $\{r$, $R\}$ & $\{\mathrm{Ur}\}$ &   & \cellcolor{C2} \num{5.42} & \cellcolor{C4} \num{9.94} & \cellcolor{C3} \num{5.54}\\
3 & $\{r$, $R$, $\mathrm{BFI}\}$ &   &   & \cellcolor{C2} \num{5.43} & \cellcolor{C4} \num{10.50} & \cellcolor{C3} \num{5.60}\\
4 & $\{r$, $R\}$ & $\{\mathrm{Ur}$, $R\}$ &   & \cellcolor{C2} \num{5.46} & \cellcolor{C4} \num{9.99} & \cellcolor{C5} \num{4.57}\\
&&&&&&\\[-1.2ex]
5 & $\{r$, $R\}$ & $\{\varepsilon$, $\widetilde{D}\}$ &   & \cellcolor{C2} \num{5.53} & \cellcolor{C3} \num{11.20} & \cellcolor{C3} \num{5.79}\\
6 & $\{r$, $\varepsilon$, $\widetilde{D}\}$ &   &   & \cellcolor{C1} \num{5.20} & \cellcolor{C3} \num{11.00} & \cellcolor{Chigh} \num{3.60}\\
7 & $\{r$, $\varepsilon$, $R\}$ &   &   & \cellcolor{C1} \num{5.31} & \cellcolor{C3} \num{11.40} & \cellcolor{C1} \num{6.97}\\
8 & $\{r$, $\widetilde{D}$, $R\}$ &   &   & \cellcolor{C2} \num{5.41} & \cellcolor{C3} \num{11.50} & \cellcolor{C0} \num{7.31}\\
&&&&&&\\[-1.2ex]
9 & $\{\varepsilon$, $\widetilde{D}$, $R\}$ &   &   & \cellcolor{Clow} \num{-0.13} & \cellcolor{Clow} \num{24.80} & \cellcolor{C0} \num{7.60}\\
10 & $\{\sigma_f\}$ & $\{\varepsilon$, $\widetilde{D}$, $R\}$ &   & \cellcolor{Clow} \num{3.93} & \cellcolor{C0} \num{13.60} & \cellcolor{Clow} \num{9.02}\\
11 & $\{r\}$ & $\{\varepsilon$, $\widetilde{D}$, $R\}$ &   & \cellcolor{C2} \num{5.41} & \cellcolor{C4} \num{10.60} & \cellcolor{C0} \num{7.18}\\
12 & $\{r\}$ & $\{\varepsilon$, $\widetilde{D}\}$ & $\{\mathrm{BFI}$, $R\}$ & \cellcolor{C2} \num{5.41} & \cellcolor{C3} \num{11.10} & \cellcolor{C2} \num{6.02}\\
&&&&&&\\[-1.2ex]
13 & $\{r$, $R\}$ & $\{\widetilde{D}$, $\varepsilon$, $\sigma_\theta\}$ &   & \cellcolor{C5} \num{5.99} & \cellcolor{C1} \num{12.40} & \cellcolor{C5} \num{4.06}\\
14 & $\{r$, $R\}$ & $\{\widetilde{D}$, $\varepsilon$, $\sigma_f\}$ &   & \cellcolor{C4} \num{5.82} & \cellcolor{C2} \num{11.80} & \cellcolor{C2} \num{6.37}\\
15 & $\{r$, $R\}$ & $\{\widetilde{D}$, $\varepsilon$, $R\}$ &   & \cellcolor{C3} \num{5.62} & \cellcolor{C3} \num{11.00} & \cellcolor{C3} \num{5.45}\\
16 & $\{r$, $\varepsilon$, $\widetilde{D}$, $\sigma_\theta\}$ &   &   & \cellcolor{C4} \num{5.83} & \cellcolor{C2} \num{11.60} & \cellcolor{C2} \num{5.94}\\
&&&&&&\\[-1.2ex]
17 & $\{r\}$ & $\{\varepsilon$, $\widetilde{D}\}$ & $\{\mathrm{BFI}$, $\sigma_f$, $\sigma_\theta\}$ & \cellcolor{C4} \num{5.85} & \cellcolor{C2} \num{11.80} & \cellcolor{C2} \num{6.40}\\
\boldmath \bfseries 18 & \boldmath \bfseries $\{r\}$ & \boldmath \bfseries $\{\varepsilon$, $\widetilde{D}$, $\sigma_f$, $\sigma_\theta\}$ & \boldmath \bfseries   & \boldmath \bfseries \cellcolor{C5} \num{6.06} & \boldmath \bfseries \cellcolor{C3} \num{11.30} & \boldmath \bfseries \cellcolor{C5} \num{4.43}\\
19 & $\{r$, $\varepsilon$, $\widetilde{D}$, $R$, $\lambda_p\}$ &   &   & \cellcolor{C4} \num{5.86} & \cellcolor{C0} \num{13.50} & \cellcolor{C0} \num{7.13}\\
20 & $\{r$, $\varepsilon$, $\widetilde{D}$, $\sigma_\theta$, $\nu\}$ &   &   & \cellcolor{Chigh} \num{6.18} & \cellcolor{C1} \num{12.80} & \cellcolor{C1} \num{6.78}\\
&&&&&&\\[-1.2ex]
21 & $\{r$, $\varepsilon$, $\widetilde{D}$, $\sigma_\theta$, $\nu$, $E_h\}$ &   &   & \cellcolor{Chigh} \num{6.19} & \cellcolor{Clow} \num{14.10} & \cellcolor{C1} \num{6.71}\\
22 & $\{r$, $\varepsilon$, $\widetilde{D}$, $\sigma_\theta$, $\sigma_f$, $\nu$, $E_h\}$ &   &   & \cellcolor{Chigh} \num{6.42} & \cellcolor{Clow} \num{15.70} & \cellcolor{C4} \num{4.97}\\
23 & $\{r$, $\varepsilon$, $\widetilde{D}$, $\sigma_\theta$, $\sigma_f$, $E_h$, $\mathrm{BFI}$, $R\}$ &   &   & \cellcolor{Chigh} \num{6.60} & \cellcolor{Clow} \num{17.00} & \cellcolor{C3} \num{5.75}\\
24 & $\{r$, $\varepsilon$, $\widetilde{D}$, $\sigma_\theta$, $\sigma_f$, $E_h$, $H_s$, $\overline{T}$, $\kappa$, $\mu$, $\lambda_p\}$ &   &   & \cellcolor{Chigh} \num{6.51} & \cellcolor{Clow} \num{19.50} & \cellcolor{C4} \num{4.95}\\
    \bottomrule
\end{tabular}
\endgroup

\vspace{2em}

\begingroup
\renewcommand{\arraystretch}{1.0}
\begin{tabular}{ll@{\hskip 3em}ll}
    \multicolumn{4}{l}{Symbols} \\
    \toprule
    $r$ & Crest-trough correlation &
    $\nu$ & Spectral bandwidth (narrowness) \\
    $\sigma_f$ & Spectral bandwidth (peakedness) &
    $\sigma_\theta$ & Directional spread \\
    $\varepsilon$ & Peak steepness $H_s k_p$ &
    $R$ & Directionality index $\sigma_\theta^2 / (2 \nu^2)$ \\
    $\mathrm{BFI}$ & Benjamin-Feir index &
    $\widetilde{D}$ & Relative peak water depth $D k_p / (2\pi)$ \\
    $E_h$ & Relative high-frequency energy &
    $\mathrm{Ur}$ & Ursell number \\
    $\overline{T}$ & Mean period &
    $\kappa$ & Kurtosis \\
    $\mu$ & Skewness &
    $H_s$ & Significant wave height \\
    \bottomrule
\end{tabular}
\endgroup

\end{table*}

\subsection*{Reference wave height distributions} We use the following theoretical wave height exceedance distributions for comparison (with rogue wave threshold $\kappa$, here $\kappa=2$):

\begin{itemize}
\item Rayleigh \citep{longuet1952statisticaldistribution}:
\begin{equation}
    P_\text{R}(\kappa) = \exp\big(-2 \kappa^2 \big)
\end{equation}

\item Tayfun \citep{tayfun1990distribution,tayfun_wave-height_2007}:
\begin{equation}
    P_\text{T}(\kappa) = \exp\bigg(\frac{-4}{1 + r} \kappa^2 \bigg) \label{eq:tayfun}
\end{equation}

\item Mori \& Janssen \citep{mori2011estimation,mori_kurtosis_2006}:
\begin{equation}
    P_\text{MJ}(\kappa) = \bigg( 1 + \frac{2\pi}{3 \sqrt{3}}  \frac{\mathrm{BFI}^2}{1 + 7.1 R} \kappa^2 (\kappa^2 - 1) \bigg) \exp\big(-2 \kappa^2 \big)
\end{equation}

\item Hybrid:
\begin{equation}
    P_\text{H}(\kappa) = \bigg( 1 + \frac{2\pi}{3 \sqrt{3}} \frac{\mathrm{BFI}^2}{1 + 7.1 R} \kappa^2 (\kappa^2 - 1) \bigg) \exp\bigg(\frac{-4}{1 + r} \kappa^2 \bigg)
\end{equation}
\end{itemize}

\subsection*{Data Availability}
The preprocessed and aggregated version of the FOWD CDIP data used in this study is available for download at \url{https://erda.ku.dk/archives/ee6b452c1907fbd48271b071c3cee10e/published-archive.html}. All model code is openly available at \url{https://github.com/dionhaefner/rogue-wave-discovery}.
}

\clearpage
\showmatmethods{} 

\acknow{Dion Häfner received funding from the Danish Offshore Technology Centre (DOTC).
Raw data were furnished by the Coastal Data Information Program (CDIP), Integrative Oceanography Division, operated by the Scripps Institution of Oceanography, under the sponsorship of the U.S. Army Corps of Engineers and the California Department of Parks and Recreation.
Computational resources were provided by DC$^3$, the Danish Center for Climate Computing.
Portions of this work were developed from the doctoral thesis of Dion Häfner \citep{hafner2022anocean}.
The authors thank Jonas Peters for helpful discussions in the early stages of this work. 
The authors thank two anonymous reviewers for their helpful comments.
This publication was made possible by the following open-source software stack: JAX \citep{jax2018github}, flax \citep{flax2020github}, optax \citep{optax2020github}, PySR \citep{cranmer2023interpretable}, scikit-learn \citep{scikit-learn}, PyALE \citep{pyale2020}, NumPy \citep{harris2020array}, SciPy \citep{2020SciPy-NMeth}, matplotlib \citep{Hunter:2007}, Seaborn \citep{Waskom2021}, pandas \citep{mckinney-proc-scipy-2010}, Jupyter \citep{jupyter}.}

\showacknow{} 

\clearpage

\bibliography{references}

\end{document}